\documentclass[12pt]{article}
\usepackage{amsmath, amsfonts,euscript,bbm}
\usepackage{epsfig, cite}
\usepackage{color}
\usepackage{ifthen}
\usepackage{graphicx}
\usepackage{setspace} 
\newboolean{Notes}\setboolean{Notes}{true}
\newcommand{\notes}[1]{\ifthenelse{\boolean{Notes}}{\textcolor{black}{#1}}{}}

\newcommand{\skyp}[1]{}

\begin{document}

\bigskip
\hskip 4in\vbox{\baselineskip12pt \hbox{FERMILAB-PUB-07-187-A} }
\bigskip\bigskip\bigskip\bigskip\bigskip\bigskip\bigskip\bigskip

\centerline{\Large Interactions of Cosmic Superstrings}
\bigskip
\bigskip
\bigskip
\centerline{\bf Mark G. Jackson}
\medskip
\centerline{Particle Astrophysics Center}
\centerline{Fermi National Accelerator Laboratory}
\centerline{Batavia, IL 60510}
\centerline{\it markj@fnal.gov}
\bigskip
\bigskip
\bigskip
\bigskip

\begin{abstract}
We develop methods by which cosmic superstring interactions can be studied in detail.  These include the reconnection probability and emission of radiation such as gravitons or small string loops.   Loop corrections to these are discussed, as well as relationships to $(p,q)$-strings. These tools should allow a  phenomenological study of string models in anticipation of upcoming experiments sensitive to cosmic string radiation.
\end{abstract}

\newpage            
\baselineskip=18pt

%%%%%%%%%%%%%%%%%%%%%%%%%%%%%%%%%%%%%%%%%%%%%%%%%%%%%%%%%%%%%%%%%%%%%%%%%%%%%%%%
\section{Introduction}
%%%%%%%%%%%%%%%%%%%%%%%%%%%%%%%%%%%%%%%%%%%%%%%%%%%%%%%%%%%%%%%%%%%%%%%%%%%%%%%%
The idea that cosmic strings might have formed in the early universe has been around for some time, being a generic consequence of $U(1)$ symmetry breaking.  If observed, these long filaments of energy stretched across the sky would be the highest energy objects ever seen.  Even more spectacular is the idea that these cosmic strings might be cosmic superstrings.  This idea was first proposed by Witten \cite{Witten:1985fp}, but for several technical reasons this was found to be unfeasible.  Progress in superstring theory, particularly non-perturbative aspects, allowed the subject to be revisited recently \cite{Copeland:2003bj} \cite{Becker:2005pv} with encouraging results.  For more complete reviews of the subject see \cite{Polchinski:2004ia} \cite{Davis:2005dd}.  

Since it is now at least plausible that cosmic strings might be observed, it is important to know how one might differentiate conventional cosmic strings from cosmic superstrings.  The former are classical objects, being an effective description of a field theory vortex solution.  Superstrings, however cosmically extended they may be, are inherently quantum objects.  Ideally these quantum fluctuations would provide observable differences in the cosmic string's behavior, allowing us to determine which type of string it is.  The issue is doubly important since experimental evidence of string theory from colliders is not expected to be forthcoming in the near future, and this may be the best opportunity to prove string theory is the correct theory of nature.

Wound states appear in the perturbative spectrum of the bosonic, type II and heterotic string theories but there has been relatively little investigation into their interactions.  The first was by Polchinski and Dai \cite{Polchinski:1988cn} \cite{Dai:1989cp} who used the optical theorem to compute the reconnection probability for bosonic wound strings.  Such scattering was also studied by Khuri \cite{Khuri:1993ax} finding that interaction is suppressed in the large-winding limit, whereas Mende used path integral saddle points \cite{Mende:1994wf} to show that for some special configurations it may still occur.  Reconnection probabilities of wound superstrings were studied in \cite{Jackson:2004zg}, and further investigation into the effect of backgrounds was performed in \cite{Jackson:2006rb}.  

In this article we develop new methods for calculating cosmic superstring interactions.  We first review the reconnection process, and show explicitly that the probability of reconnection can be obtained by summing over all final kinked states.  This method naturally generalizes to the possibility of emitting radiation during reconnection, and allows us to calculate the probability for this as well.  We discuss the spacetime trajectory of the strings during reconnection, as well as loop corrections to the interactions.  A novel boundary functional representation of the kinked string is also given.  Finally we conclude with an overview of upcoming experiments which might be sensitive to these signatures.  To emphasize the principles introduced here we study only the bosonic string, but the ideas should easily generalize to the superstring.

\begin{figure}
\begin{center}
\includegraphics{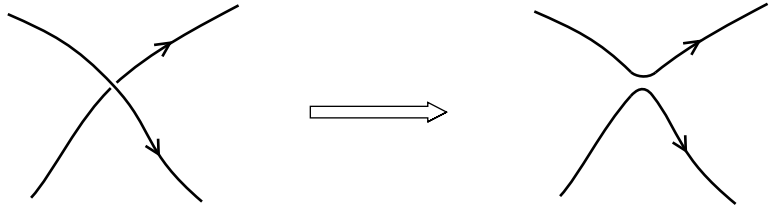}
\caption{When cosmic strings approach each other, there is some probability for them to reconnect.} \ \\
\end{center}
\end{figure}

%%%%%%%%%%%%%%%%%%%%%%%%%%%%%%%%%%%%%%%%%%%%%%%%%%%%%%%%%%%%%%%%%%%%%%%%%%%%%%%%
\section{The Cosmic Superstring Vertex Operators and States}
%%%%%%%%%%%%%%%%%%%%%%%%%%%%%%%%%%%%%%%%%%%%%%%%%%%%%%%%%%%%%%%%%%%%%%%%%%%%%%%%
We will model cosmic superstrings as wound fundamental string states.  These will then interact to form other wound states, but will typically be kinked due to the relative angle between the initial states (see Figure 1).  The most straightforward way to study these interactions is to construct vertex operators for all states.  Consider two long, straight wound strings on a 2D torus of size $L$ and skew angle $\theta$ as illustrated in Figure 2.  The momenta are taken to be 
\begin{figure}
\begin{center}
\includegraphics{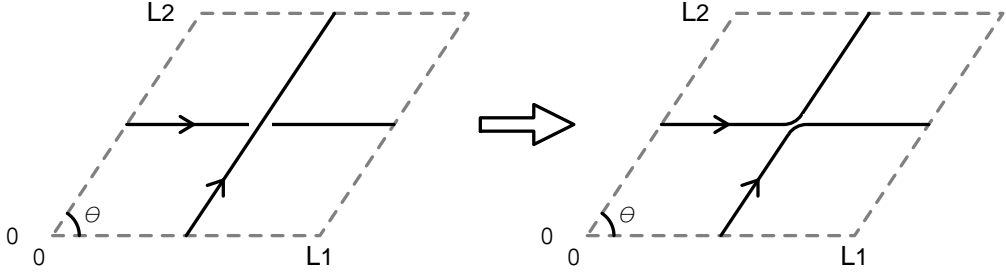}
\caption{We model cosmic superstrings as straight wound modes on a large torus, which will then interact to form a kinked configuration.} \ \\
\end{center}
\end{figure}
\[ p_1 = \left[ \left( \frac{L}{2 \pi \alpha'} \right)^2 - \frac{4}{\alpha'} \right]^{1/2} (1,0,0,0, {\bf 0} ), \hspace{0.5in} L_1 = L (0,1,0,0, {\bf 0}), \]
\[ p_2 = \left[ \left( \frac{L}{2 \pi \alpha'} \right)^2 - \frac{4}{\alpha'} \right]^{1/2} [1-v^2]^{-1/2} (1,0,0,v, {\bf 0} ), \hspace{0.5in} L_2 = L (0,\cos \theta,\sin \theta,0, {\bf 0}) . \]
These satisfy the conditions such there are no complications involving the branch cuts of the vertex operators (see section 8.2 of \cite{joebook}), and also the tachyonic mass-shell conditions
\[ p^2_{iL} = p^2_{iR} = \frac{4}{\alpha'}, \hspace{0.5in} p_{iL/R} = p_i \pm \frac{L_i}{2 \pi \alpha'} . \]
The relevant vertex operators are: ($i=1,2$)
\begin{equation}
\label{simplevo}
 V_T(z, {\bar z}; p_i) = \frac{\kappa}{2 \pi \sqrt{V}} :e^{ip_{iL} X_L(z) + ip_{iR} X_R ({\bar z})}:
 \end{equation}
where the volume $V = V_\perp L^2 \sin \theta $ is the product of the the transverse volume and the 2D torus (methods to calculate $V_\perp$ can be found in \cite{Jackson:2006rb}).  Now examining the OPE of these vertex operators (we will only consider the holomorphic side, the antiholomorphic side is identical):
\begin{eqnarray*}
: e^{ip_{1L} X_L(z)}::e^{ip_{2L} X_L(0)}: &=& z^{ {\alpha' \over 2} p_{1L} \cdot p_{2L} } : e^{ip_{1L}X_L(z)+i p_{2L} X_L(0)}: \\
&=& z^{ {\alpha' \over 2} p_{1L} \cdot p_{2L} }: \left( 1 + i z p_{1L} \cdot \partial X_L(0) + \ldots \right) e^{i(p_{1L}+p_{2L})X_L(0)}: .
\end{eqnarray*}
The Taylor expansion of the exponential shows the vertex operators of the infinite tower of the produced states, which will appear kinked due to their large oscillator excitation number $N$:
\begin{eqnarray*}
N-1 &=& - \frac{\alpha'}{4} (p_{1L}+p_{2L})^2 \\
&=& -2 - \frac{\alpha'}{2} p_{1L} \cdot p_{2L} \\
&\sim& L^2 / \alpha' .
\end{eqnarray*}
Since $p_{1R} \cdot p_{2R} = p_{1L} \cdot p_{2L}$ the result will be identical for the right-moving oscillators and so ${\tilde N}=N$.  The vertex operators for the possible final states (labeled by index $f$) $V_{{\rm kink},f}$ will be the $z^N$ coefficient in the expansion of the exponential, each weighted by a coefficient $\mathcal M_f$.  To extract these we simply take the contour integral in independent variables $\epsilon$ and ${\bar \epsilon}$ around the origin:
\begin{equation}
\label{kinkvertex}
\sum_f \mathcal M_f V_{{\rm kink},f}(0; p_1+p_2) =  C_{S_2} \left( \frac{\kappa}{2 \pi \sqrt{V}} \right)^2 \frac{1}{(2 \pi i)^2} \oint_0 d \epsilon d {\bar \epsilon} \ V_T(\epsilon,{\bar \epsilon};p_1) V_T(0;p_2)
\end{equation}
with $C_{S_2} = 32 \pi^3 V / \kappa^2 \alpha'$ the normalization for sphere amplitudes.  We have normalized the RHS of (\ref{kinkvertex}) so that these coefficients $\mathcal M_f$ are none other than the invariant amplitude\footnote{Note that since we have compactified 2 of the 4 Minkowski dimensions, all scattering must be considered 2-dimensional.} to produce that final kinked state,
\[ \langle V_{{\rm kink},f}(\infty; -p_f) V_T(1;p_1) V_T(0;p_2) \rangle = i (2 \pi)^2 \delta^2 (p_1 + p_2 - p_f) \mathcal M_f . \]
The number of possible final states is then that of a string excited along a single dimension $p_1$, 
\begin{equation}
\label{deg}
 D(N) \sim N^{-1} e^{2 \pi \sqrt{N/6}}. 
 \end{equation}
Owing to this large degeneracy as $N \rightarrow \infty$ in the cosmic string limit, it will be impossible to explicitly write down this sum of kinked vertex operators, but there is a simple statistical distribution.  Consider the state corresponding to the sum of these vertex operators (\ref{kinkvertex}):
\begin{eqnarray}
\label{kinkstate}
| {\rm kinks} \rangle &=& \frac{\alpha' }{8 \pi}  \left( \frac{\kappa}{2 \pi \sqrt{V}} \right)^{-1}  \sum_f \mathcal M_{f} | {\rm kink}_{f} \rangle \\
\nonumber
&=& \frac{1}{(2 \pi i)^2}  \oint_0 d \epsilon d {\bar \epsilon} \ |\epsilon|^{-2(N+1)} e^{ \sqrt{\alpha' \over 2} \sum_{n \geq 1} p_{1L} \cdot \alpha_{-n} \epsilon^n / n+ p_{1R} \cdot {\tilde \alpha}_{-n} {\bar \epsilon}^n / n} |p_{1}+p_{2} \rangle. 
\end{eqnarray}
The expectation value of the number operator $N_n = \frac{1}{n} \alpha_{-n} \alpha_{n}$ in this state is:
\begin{eqnarray*}
\frac{\langle {\rm kinks} | N_n | {\rm kinks} \rangle }{\langle {\rm kinks} |{\rm kinks} \rangle } &=& 2 \left( \frac{1}{n} - \frac{1}{N+1} \right), \hspace{0.5in} n \leq N.
\end{eqnarray*}
To evaluate the contours we have transformed the outgoing state contour variable $\epsilon \rightarrow 1/\epsilon$ so that it is also taken around the origin, and then used the standard representation
\[ (1-x)^{-n} = \frac{1}{\Gamma(n)} \int _0 ^\infty dt \ t^{n-1} e^{-t(1-x)}. \]
Although this is the spectrum expected for a kinked string, the distribution does not converge to this mean state when the excitation number grows large, as can be seen by the relative fluctuation:
\[ \frac{\langle (N_n-\langle N_n \rangle)^2 \rangle}{\langle N_n \rangle^2}  \rightarrow n. \]
It would be interesting to see whether there is any relation of this state to the kinky strings studied by McLoughlin \emph{et al.} \cite{McLoughlin:2006tz}.
%%%%%%%%%%%%%%%%%%%%%%%%%%%%%%%%%%%%%%%%%%%%%%%%%%%%%%%%%%%%%%%%%%%%%%%%%%%%%%%%
\section{Reconnection}
\subsection{Tree-Level}
%%%%%%%%%%%%%%%%%%%%%%%%%%%%%%%%%%%%%%%%%%%%%%%%%%%%%%%%%%%%%%%%%%%%%%%%%%%%%%%%
We can now use the expression (\ref{kinkvertex}) to calculate the reconnection probability that two straight strings would scatter into any final kink state, which is the interaction cross-section as shown in Figure 3:
\begin{equation}
\label{p0}
P = \frac{1}{4 E_1 E_2 v} \int \frac{d {\bf p}_{f}}{2 \pi} \frac{1}{2E_f} \sum_f |\mathcal M_f|^2 (2 \pi )^2 \delta^2(p_1 + p_2 - p_f).
\end{equation}
\begin{figure}
\begin{center}
\includegraphics[width=4in]{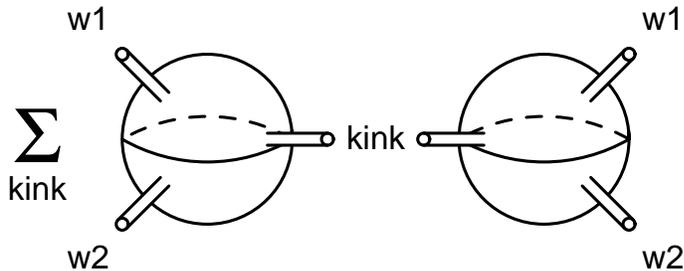}
\caption{The cosmic superstring reconnection probability is found by summing the scattering amplitude over all final kinked configurations.}
\end{center}
\end{figure}
To evaluate this we will use the orthonormal relation for two-point correlators on the sphere (we assume that both $V_i(p_i)$ and $V_j(p_j)$ are on-shell vertex operators):
\[ \langle V_i(\infty; -p_i) V_j (0; p_j) \rangle = \frac{8 \pi}{\alpha'} (2 \pi)^2 \delta^2(p_i-p_j) \delta_{ij}. \]
We also wish to transform the integral over phase space into a more useful form.  Recall that this is the space of all on-shell final states,
\[  \sum_f \int \frac{d {\bf p}_{f}}{2 \pi} \frac{1}{2E_f} = \sum_f \int \frac{d^2 p_{f}}{(2 \pi)^2} \ 2 \pi \delta(p_f^2 - m_f^2) . \]
This delta function represents the change in phase space with respect to the 2-momentum invariant $p_f^2$, which gives poles at ${\alpha' \over 4} m_f^2 \sim L^2 / \alpha'  = -1, 0, 1, 2, \ldots $.  In the large-mass limit of our cosmic strings we can perform an averaging over these poles,
\[ \sum_f \delta(p_f^2 - m_f^2) \rightarrow \frac{\alpha'}{4}. \]
 This is same averaging of poles into a branch cut performed in \cite{Polchinski:1988cn} \cite{Jackson:2004zg} in order to utilize the optical theorem.  The probability can now be easily calculated:
\begin{eqnarray}
\nonumber
 && \int \frac{d {\bf p}_{f}}{2 \pi} \frac{1}{2E_f} \sum_f |\mathcal M_f|^2 (2 \pi )^2 \delta^2(p_1 + p_2 - p_f) \\
 \nonumber
 && \hspace{0.5in} =\frac{\alpha'^2}{16} \int \frac{d^2 p_{f}}{(2 \pi)^2} \int \frac{d^2 p_j}{(2 \pi)^2} \langle \sum_f \mathcal M^*_f V_{{\rm kink},f}(\infty; -p_f) \sum_j \mathcal M_j V_{{\rm kink},j}(0; p_j) \rangle (2 \pi )^2 \delta^2(p_1 + p_2 - p_f) \\
\nonumber
 && \hspace{0.5in} = \frac{(C_{S_2} \alpha' )^2}{16} \left( \frac{\kappa}{2 \pi \sqrt{V}} \right)^4 \frac{1}{(2 \pi i)^4} \langle \oint_\infty d ^2\eta \ V_T(\infty;-p_2) V_T(\eta,-p_1) \oint_0 d^2 \epsilon \ V_T(\epsilon,p_1) V_T(0,p_2) \rangle \\
\nonumber
 && \hspace{0.5in} = \frac{C_{S_2}^3 \alpha'^2}{16}  \left( \frac{\kappa}{2 \pi \sqrt{V}} \right)^8 \frac{1}{(2 \pi i)^4} \left[ \oint_0 d \eta d\epsilon \ (\epsilon \eta)^{-(N+1)} (1-\epsilon \eta)^{-2} \right]^2 \\
\label{ampsum}
 && \hspace{0.5in} = \frac{8 \pi \kappa^2}{{\alpha'} V} (N+1)^2.
\end{eqnarray} 
Substituting back for $N+1=-{\alpha' \over 2} p_{1L} \cdot p_{2L}$, taking the $L \rightarrow \infty$ limit, and writing the answer in terms of the dimensionless coupling $g_s$, the probability of reconnection is simply
\begin{equation}
\label{p0}
P = g_s^2 \frac{V_{\rm min} }{V_\perp}  \frac{ (1- \cos \theta \sqrt{1-v^2})^2}{8 \sin \theta v \sqrt{1-v^2}},  \hspace{0.5in} V_{\rm min} = (4 \pi^2 \alpha')^3 . 
\end{equation}
%%%%%%%%%%%%%%%%%%%%%%%%%%%%%%%%%%%%%%%%%%%%%%%%%%%%%%%%%%%%%%%%%%%%%%%%%%%%%%%%
\subsection{A Better Way to Do This}
%%%%%%%%%%%%%%%%%%%%%%%%%%%%%%%%%%%%%%%%%%%%%%%%%%%%%%%%%%%%%%%%%%%%%%%%%%%%%%%%
Let us now rederive the results in the previous section, using an abbreviated notation which will be very useful for more complicated interactions.

Just as we used the contour integral to extract the on-shell part of the ${: e^{ip_{1} X(z,{\bar z})}:}:e^{ip_{2} X(0)}:$ OPE, the $| \rm{ kinks} \rangle$ state will be the part of $: e^{ip_{1} X(z,{\bar z})}:|p_2 \rangle$ that it is annihilated by both $L_0 -1$ and ${\tilde L}_0 -1$:
\[ (L_0 -1) | {\rm kinks} \rangle = ({\tilde L}_0 -1) | \rm{ kinks} \rangle = 0 .  \]
A projection operator which accomplishes this can easily be constructed to be
\begin{eqnarray*}
\mathcal P &=&  \frac{ \sin \pi (L_0 -1)}{\pi (L_0 -1)} \frac{ \sin \pi ({\tilde L}_0 -1)}{\pi ({\tilde L}_0 -1)} \\
&=& \frac{1}{2 \pi} \int _{-\pi} ^\pi d \sigma_L \ e^{i \sigma_L (L_0-1)} \ \frac{1}{2 \pi} \int _{-\pi} ^\pi d \sigma_R \ e^{i \sigma_R ({\tilde L}_0-1) }
\end{eqnarray*}
so that
\begin{eqnarray}
\label{holokink}
 | {\rm kinks} \rangle  &=&  \left( \frac{\kappa}{2 \pi \sqrt{V}} \right)^{-1}  \mathcal P V_T(1;p_1) |p_2 \rangle \\
 \nonumber
 &=&   \left( \frac{\kappa}{2 \pi \sqrt{V}} \right)^{-1} \frac{1}{(2 \pi)^2}  \int _{-\pi} ^\pi d \sigma_L d \sigma_R \ V_T(e^{i \sigma_L},e^{-i \sigma_R};p_1) |p_2 \rangle.
 \end{eqnarray}
Identifying $e^{i\sigma_L} \rightarrow \epsilon, e^{-i\sigma_R} \rightarrow {\bar \epsilon}$ then yields the previous expression (\ref{kinkstate}).  The sum over final states can now be written as simply
\[ \int \frac{d {\bf p}_{f}}{2 \pi} \frac{1}{2E_f} \sum_f |\mathcal M_f|^2 (2 \pi )^2 \delta^2(p_1 + p_2 - p_f) = \frac{32 \pi ^3}{\alpha'} \langle p_2 | V_T(-p_1) \mathcal P V_T(p_1)|p_2 \rangle. \]
We can use merely a single $\mathcal P$ since like other projection operators ${\mathcal P}^2 = \mathcal P$.  It is then straightforward to see that this yields (\ref{ampsum}).  This formalism also allows us to make contact with the use of the optical theorem \cite{Polchinski:1988cn} \cite{Jackson:2004zg}, written
\[  \int \frac{d {\bf p}_{f}}{2 \pi} \frac{1}{2E_f} \sum_f |\mathcal M_f|^2 (2 \pi )^2 \delta^2(p_1 + p_2 - p_f) = \left( \frac{8 \pi}{\alpha'} \right)^2 {\rm Im} \ \langle p_2 | V_T(-p_1) \Delta V_T(p_1)|p_2 \rangle \]
where $\Delta$ is the closed string propagator with Feynman $i \epsilon$ prescription,
\[ \Delta = \frac{\alpha' \delta(L_0-{\tilde L}_0)}{2(L_0+{\tilde L}_0-2-i \epsilon)}. \]
Let us now trade the (anti)holomorphic operators $L_0, {\tilde L}_0$ for the worldsheet hamiltonian $H$ and momentum $P$ operators:
\begin{eqnarray*} 
H &=& L_0 + {\tilde L}_0 - 2, \\
P &=& L_0 - {\tilde L}_0.
\end{eqnarray*}
The propagator can then be written as
\begin{eqnarray*}
{\rm Im} \ \Delta &=& \frac{\alpha' }{4 i}\delta(P)  \left( \frac{1}{H-i \epsilon} - \frac{1}{H+i \epsilon} \right) \\
 &=& \frac{\pi \alpha'}{2} \delta(P) \delta (H) \\
 &=& \frac{ \pi \alpha'}{4} \mathcal P.
 \end{eqnarray*}
 This is then identical to the expression given above.
%%%%%%%%%%%%%%%%%%%%%%%%%%%%%%%%%%%%%%%%%%%%%%%%%%%%%%%%%%%%%%%%%%%%%%%%%%%%%%%%
\subsection{Intercommutation with Radiation Emission }
%%%%%%%%%%%%%%%%%%%%%%%%%%%%%%%%%%%%%%%%%%%%%%%%%%%%%%%%%%%%%%%%%%%%%%%%%%%%%%%%
The probability for intercommutation should also include the possibility that radiation could be emitted in the process of reconnection.  Such emission has been studied previously using other techniques \cite{Cornou:2006wx} \cite{Melkumova:2006ua} but here we show it is a simple generalization of the same process used to compute ordinary reconnection probabilities.  The amplitude that a single state $V_{\rm rad}(k)$ will be emitted during the reconnection to kink $f$ is given by
\begin{equation}
\label{mrad}
 \mathcal A_{{\rm rad},f}(k) = \int d^2 z \ \langle V_{{\rm kink}',f}(\infty;-p_1-p_2+k) V_{\rm rad}(z;-k) V_T(1;p_1) V_T(0;p_2) \rangle.
 \end{equation}
 \begin{figure}
\begin{center}
\includegraphics[width=6in]{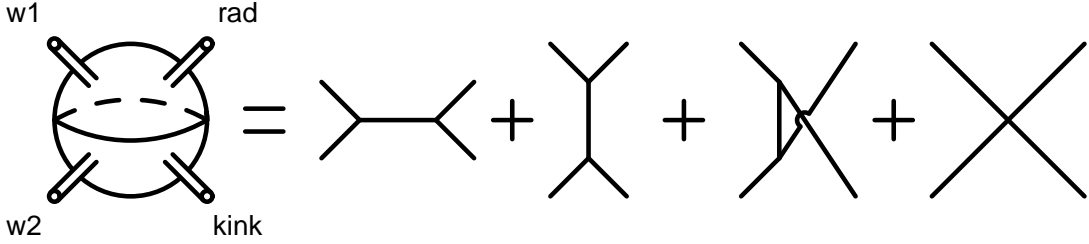}
\caption{The cosmic superstring reconnection process while emitting radiation represents four diagrams in field theory.  Only the first and last are found to be important in the cosmic string limit.}
\end{center}
\end{figure}
 This amplitude represents four different field theory processes: the $s$-, $t$-, and $u$-channel interactions as well as a contact term as shown in Figure 4.  Actually, only the first and last of these contribute.  Consider the excitation number of the intermediate states: for the $s$-channel this is the same excitation number as for the kink computed previously,
\begin{eqnarray*}
N_s-1 &=& - \frac{\alpha'}{4} (p_{1L}+p_{2L})^2 \\
&\sim& L^2 / \alpha'
\end{eqnarray*}
whereas for $t$- and $u$- this is
\begin{eqnarray*}
N_{t,u}-1 &=& - \frac{\alpha'}{4} (p_{iL}-k)^2 \\
&=& -1 +  \frac{\alpha'}{4} m^2_{\rm rad} + \frac{\alpha'}{2} p_{iL} \cdot k \\
&\sim& -L / \sqrt{\alpha'}.
\end{eqnarray*}
Since we cannot have a negatively excited string, there are no intermediate states in these channels.

Nonetheless, we consider the expression representing all four channels by writing the expression for the sum over (perturbed) kink states in terms of the two distinct vertex operator orderings, where we integrate in $w$ to include the pole over the intermediate states (rather than contour integrate to simply get the residue):
 \begin{eqnarray*}
&& \hspace{-0.7in} \sum_f \mathcal M_{{\rm rad},f}(k)V_{{\rm kink'},f}(0; p_1+ p_2-k) = \\
 && \hspace{-0.4in}  \frac{C_{S_2} \alpha'}{4 \pi} \left( \frac{\kappa}{2 \pi \sqrt{V}} \right)^2 \frac{1}{(2 \pi i)^2} \oint_0 d \epsilon d {\bar \epsilon}  \int_{|w|<1} d^2 w \ V_{\rm rad}(\epsilon; -k) V_T(w; p_1) V_T(0; p_2) + (V_{\rm rad} (-k)\leftrightarrow V_T(p_1) ), \\
 | {\rm kinks}' \rangle &=& \frac{8 \pi}{\alpha'}   \mathcal P \left( V_{\rm rad}(-k) \Delta V_T(p_1) |p_2 \rangle + V_T(p_1) \Delta V_{\rm rad}(-k) |p_2 \rangle \right).
\end{eqnarray*}
With the new possibility that the radiation could be directed into the $x$-$y$ plane, there is the slight complication that we have compactified along these directions so the momentum must be discrete rather than continuous.  As appropriate in the large-$L$ limit, we approximate the sum over KK modes as an integral over continuous momenta times the compactification volume,
\[ \sum_k \rightarrow L^2 \sin \theta \int \frac{d^3 {\bf k}}{(2 \pi)^2} \frac{1}{2 E_{\bf k}} . \] 
We also add a 2D sum over $p_f$ and momentum-conservation integral, for which the volume factors cancel,
\[ \sum_{p_f,xy} \delta^2_{k+p_f,xy} \rightarrow  \int \frac{d^2{\bf p}_{f,xy}}{(2 \pi)^2} \ (2 \pi)^2 \delta^2( {\bf k}_{xy}+{\bf p}_{f,xy} ). \]
Making these modifications to the reconnection probability given in (\ref{p0}), we arrive at
\begin{equation}
\label{prad}
 P_{\rm rad} = \frac{L^2 \sin \theta}{4 E_1 E_2 v} \int \frac{d ^3 {\bf p}_f}{(2 \pi)^3} \frac{1}{2E_f} \frac{d ^3 {\bf k}}{(2 \pi)^3} \frac{1}{2E_{\bf k}} \sum_f |\mathcal M_{{\rm rad},f}(k)|^2 (2 \pi )^4 \delta^4(p_1 + p_2 -k- p_f).
 \end{equation}
This is evaluated using the same techniques as in the simple reconnection case, averaging the poles of $p_f$ (but \emph{not} k), and multiplying by an additional $(2 \pi \sqrt{ \alpha'})^{-2}$ for the units introduced by the propagators,
\begin{eqnarray*}
 && \hspace{-0.8in} \int \frac{d ^3 {\bf p}_f}{(2 \pi)^3} \frac{1}{2E_f} \sum_f |\mathcal M_{{\rm rad},f}(k)|^2 (2 \pi )^4 \delta^4(p_1 + p_2 -k- p_f)  \\
\ &=&\frac{8 \pi}{\alpha'^2}   \langle p_2 | V_T(-p_1) \Delta V_{\rm rad}(k) \mathcal P V_{\rm rad}(-k) \Delta V_T(p_1) |p_2 \rangle + {\rm perms}.
\end{eqnarray*}
Using the integral representation of the propagator,
\[ \Delta = \frac{\alpha'}{4 \pi} \int_{|z| <1} \frac{d^2 z}{|z|^2} z^{L_0-1} {\bar z}^{{\tilde L}_0 -1} \]
we can act on the radiation vertex operators in every permutation, since $V_{\rm rad}$ will always be adjacent to either $|p_2 \rangle$ or $\mathcal P$.  For example, 
\begin{eqnarray*}
z^{L_0-1} V_{\rm rad}(1) |p_2 \rangle &=& z^{L_0-1} V_{\rm rad}(1) z^{-(L_0-1)} |p_2 \rangle  \\
&=& V_{\rm rad}(z) |p_2 \rangle.
\end{eqnarray*}
Adding all permutations together (corresponding to different regions of integration for the radiation), this results in integration over the entire complex plane.  We then interpret the reconnection process as a background for the radiation, meaning we neglect the cross-radiation term\footnote{Besides the intuition that such cross-radiation terms should be negligible, for more realistic relativistic radiation we would have ${\alpha' \over 2} k^2 \approx 0$, and the correlation between radiation vertex operators disappears.} as well as the effect that the radiation would have on the projection onto physical states.  The resultant sum over final states is then simply (neglecting phase-space factors)
\begin{equation}
\sum_f | \mathcal M_{{\rm rad},f} |^2 = \frac{8 \pi}{\alpha'^2} \langle p_2 | V_T(-p_1) \mathcal P V_T(p_1)|p_2 \rangle \left| \int d^2 z \ V_{\rm rad} [k;X_{\rm cl}(z, {\bar z})] \right|^2 
\end{equation}
where the classical position $X_{\rm cl}$ is defined as the mean value during the reconnection,
\begin{eqnarray}
\nonumber
X_{\rm cl} (z,{\bar z}) &=& \frac{ \langle p_2 | V_T(-p_1) \mathcal P X(z,{\bar z}) V_T(p_1)|p_2 \rangle }{\langle p_2 | V_T(-p_1) \mathcal P V_T(p_1)|p_2 \rangle } \\
\nonumber
&=& i \left. \frac{\partial}{\partial k} \right|_{k=0} \ln \ \langle p_2 | V_T(-p_1) \mathcal P :e^{-ikX(z,{\bar z}) }: V_T(p_1)|p_2 \rangle\\
\nonumber
&=&  i \left. \frac{\partial}{\partial k} \right|_{k=0} \ln \left[ z^{ -{\alpha' \over 2} k \cdot p_{2L} } (z-1)^{ -{\alpha' \over 2} k \cdot p_{1L} } \oint _0 d \epsilon \ \epsilon^{-(N+1)} (1-\epsilon)^{-2} (1 - \epsilon z)^{ {\alpha' \over 2} k \cdot p_{1L} } \times (L \rightarrow R) \right] \\
\label{spacetime}
&=&  -i {\alpha' \over 2} p_{2L} \ln z -i {\alpha' \over 2} p_{1L} \left[ \ln (z-1) + \sum _{n=1}^N \left( \frac{1}{n} - \frac{1}{N+1} \right) z^n \right] + (L \rightarrow R). \hspace{0.4in}
 \end{eqnarray}
The straight-string vertex operators produce the terms $-i {\alpha' \over 2} p_{2L} \ln z$ and $ -i {\alpha' \over 2} p_{1L} \ln (z-1)$, which have the expected branch cut on the real axis for $z \leq 0,1$ representing the windings $L_2,L_1$, respectively, as one crosses the cuts.  In the large-winding limit we see there is an additional branch cut produced on this axis for $z \geq 1$,
\begin{equation}
\label{largenxcl}
 X_{\rm cl}(z,{\bar z}) \rightarrow  -i {\alpha' \over 2} p_{2L} \ln z -i  {\alpha' \over 2} p_{1L} \ln \left( \frac{z-1}{1-z} \right) + (L \rightarrow R) \ {\rm as} \ N \rightarrow \infty.
 \end{equation}
As one crosses this at large $z$, the two branch cuts conspire to replace the gradual winding on $L_1$ with a step-function.
  
The probability (\ref{prad}) can now be written as the simple probability $P_0$ times a radiative phase-space factor:
 \begin{eqnarray}
 \nonumber
 P_{\rm rad} &=& \frac{L^2 \sin \theta}{4 E_1 E_2 v} \frac{8 \pi \kappa^2}{\alpha' V} (N+1)^2 \left( \frac{\kappa}{2 \pi \sqrt{V}} \right)^2 \frac{\alpha'}{(4 \pi)^4} \int \frac{d ^3 {\bf k}}{(2 \pi)^3} \frac{1}{2E_{\bf k}}\left| \int d^2 z \ V_{\rm rad} [k;X_{\rm cl}(z, {\bar z})] \right|^2 \\
 \label{prad2}
 &=& P_0 \left( \frac{\alpha' g^2_s V_{\rm min}}{(4 \pi)^4 V_\perp } \right)  \int \frac{d ^3 {\bf k}}{(2 \pi)^3} \frac{1}{2E_{\bf k}} \left| \int d^2 z \ V_{\rm rad} [k;X_{\rm cl}(z, {\bar z})] \right|^2 .
 \end{eqnarray} 
This partly illuminates a puzzle discovered in \cite{Jackson:2004zg}, that there is a $1/v$ divergence in the fundamental string reconnection probability.  Since the integrand in (\ref{prad2}) universally contains the exponential of $p_{2z} k_z \propto v k_z $, the integration over $k_z$ will result in an additional factor of $1/v$ (recall the original $1/v$ originated from the incoming strings' kinematic factor).  Again taking the cross-radiation effects to be negligible, the result immediately generalizes to multiple radiative emissions:
\[ P_{{\rm rad},n} = P_0 \left( \frac{\alpha' g^2_s V_{\rm min}}{(4 \pi)^4 V_\perp } \right)^n \prod_{i=1}^n \int \frac{d ^3 {\bf k}_i}{(2 \pi)^3} \frac{1}{2E_{{\bf k},i}} \left| \int d^2 z_i \ V_{\rm rad} [k_i;X_{\rm cl}(z_i, {\bar z_i})] \right|^2 . \]
The probability that there are $n$ states radiated during reconnection will have $n$ such radiated momenta integrals and produce a factor of $1/v^n$.  The tree-level reconnection probability can then be put in the form 
\[ P = \sum_{n=0} ^\infty P_n \left( \frac{g_s^{2}}{v} \right)^{n+1}. \]
This implies that the effective coupling seen by the strings is actually $\lambda \equiv g_s^2/v$ so that the perturbative, small velocity limit should be defined by $g_s^2 \rightarrow 0, v \rightarrow 0$ and fixed $\lambda$.  Loop corrections to this simply add factors of $g_s^2$ which vanish, so in some sense the strings become classical in this limit.

The resultant integral of the radiation vertex operator over $z$ is then easily evaluated since as explained in \cite{Damour:2000wa} the radiation amplitude drops exponentially with energy unless there exist double saddle points $k \cdot \partial X(z_0)=k \cdot {\bar \partial} X(z^*_0) = 0$ (such as for a cusp) in which case it is a slower power-law decay or when the (say) holomorphic side has a saddle point and the antiholomorphic derivative is discontinuous (such as for a kink), which is what happens in (\ref{largenxcl}).  The radiation properties of the reconnection are currently being investigated in more detail \cite{mgj2}.

 %%%%%%%%%%%%%%%%%%%%%%%%%%%%%%%%%%%%%%%%%%%%%%%%%%%%%%%%%%%%%%%%%%%%%%%%%%%%%%%%
\section{Spacetime Picture of String Reconnection}
%%%%%%%%%%%%%%%%%%%%%%%%%%%%%%%%%%%%%%%%%%%%%%%%%%%%%%%%%%%%%%%%%%%%%%%%%%%%%%%%
\begin{figure}
\begin{center}
\includegraphics[width=3in]{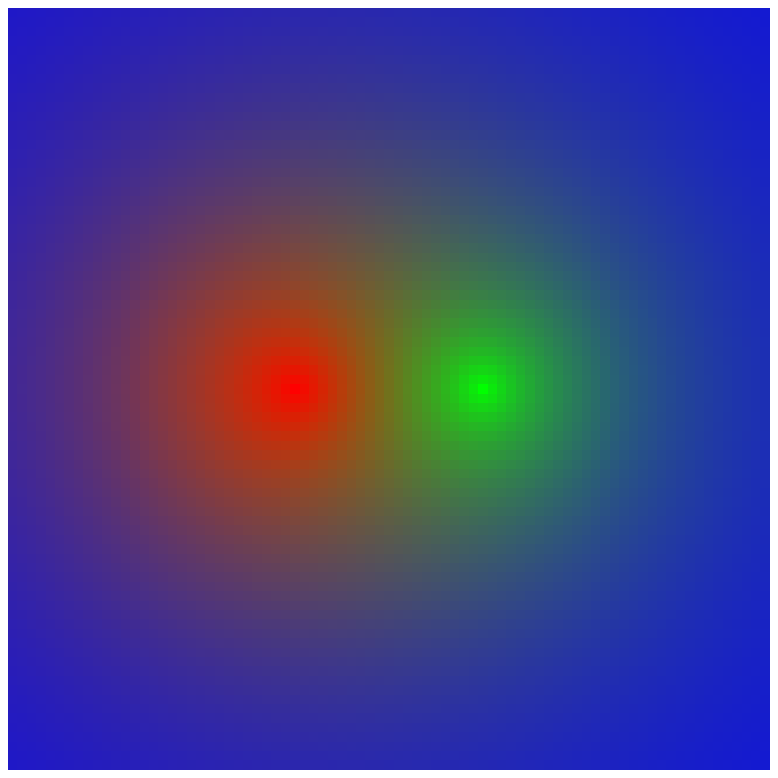}
\caption{The reconnection process from the worldsheet viewpoint.  The red, green and blue colors represent vertex operators at $z=\{ 0,1,\infty \}$, respectively.} \ \\
\includegraphics[width=4in]{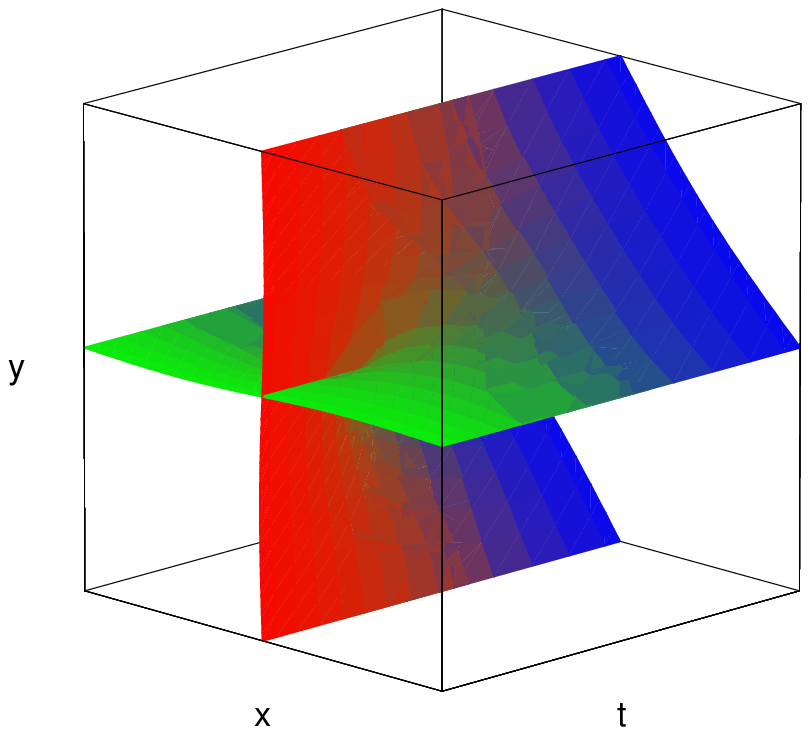}
\caption{A qualitative representation of the reconnection process from the spacetime viewpoint (the $z$-axis has been projected out).} \ \\
\end{center}
\end{figure}
Of course the classical trajectory (\ref{spacetime}) cannot literally be interpreted as the spacetime path swept out by the superstrings during reconnection, since it is not reparameterization-invariant: only the amplitude $\int d^2 z \ V_{\rm rad}[X(z, {\bar z})]$ is meaningful.  But it can give us some idea of how string theory sees the reconnection process, both from the worldsheet and spacetime viewpoints.

In Figure 5 the string worldsheet is shown with the coloring added to illustrate how these asymptotic string states smoothly reconnect: red, green and blue indicate the ``pure" states at $\{ 0,1,\infty \}$, respectively.  Good local coordinates near the straight-string vertex operators at $z_i =  \{ 0,1 \}$ are $e^{\tau + i\sigma} = z-z_i$, and for the kinked state placed at infinity we use $e^{\tau + i\sigma} = 1/z$.  In these coordinates $\sigma \rightarrow \sigma + 2\pi$ generates $X \rightarrow X + L_i$.    In the spacetime embedding of \emph{only the straight string vertex operators} $-i {\alpha' \over 2} p_{2L} \ln z$ and $ -i {\alpha' \over 2} p_{1L} \ln (z-1)$ shown in Figure 6, we see the initial straight wound strings arrive from $t \rightarrow -\infty$, then smoothly join into a single kinked string as $t \rightarrow \infty$.  Although the strings appear to break, this is merely the effect of the worldsheet embedding into spacetime and then slicing it with respect to $t$.
%%%%%%%%%%%%%%%%%%%%%%%%%%%%%%%%%%%%%%%%%%%%%%%%%%%%%%%%%%%%%%%%%%%%%%%%%%%%%%%%
\section{Loop Corrections}
%%%%%%%%%%%%%%%%%%%%%%%%%%%%%%%%%%%%%%%%%%%%%%%%%%%%%%%%%%%%%%%%%%%%%%%%%%%%%%%%
We have seen that the probability of reconnection should include not just the probability of simple reconnection but processes which produce radiation.  In fact one should also include higher-order corrections for each of these (omitting phase-space factors),
\[  P = \frac{1}{4 E_1 E_2 v} \sum_f  \left[ |\mathcal M_f^{(0)} + \mathcal M_f^{(1)} + \cdots |^2 + |\mathcal M_{{\rm rad},f}^{(0)} + \mathcal M_{{\rm rad},f}^{(1)} + \cdots |^2 + \cdots \right]  . \]
The most convenient way to compute the loop effects would be to compute the correction to the total cross-section:
\begin{equation}
\label{naivecor}
P = \frac{1}{2 E_1 E_2 v} \sum_{n \geq 0} {\rm Im} \left. \mathcal M^{(n)} \right|_{t=0}. \hspace{0.5in} ({\rm naive}) 
\end{equation}
In \cite{Jackson:2004zg} it was noted that this diverged at higher loops since $\mathcal M^{(n)} \sim L^{2+2n}$, but the cause was suspected to be the incorrect inclusion of amplitudes having multiple final kinked string states, each state having normalization $\sim L^2/\alpha'$.   The correct approach would be evaluating (\ref{naivecor}) while excising the intermediate states containing multiple kinked strings (since these would not correspond to a reconnection!).  This could be done by keeping only the part of each $\mathcal M^{(n)}$ proportional to $L^2$, corresponding to final states containing only a single kinked string.  Investigation along these lines is underway \cite{oneloop}.
%%%%%%%%%%%%%%%%%%%%%%%%%%%%%%%%%%%%%%%%%%%%%%%%%%%%%%%%%%%%%%%%%%%%%%%%%%%%%%%%
\section{Radiation}
%%%%%%%%%%%%%%%%%%%%%%%%%%%%%%%%%%%%%%%%%%%%%%%%%%%%%%%%%%%%%%%%%%%%%%%%%%%%%%%%
Now let us consider radiation not from reconnection, but simply emitted from a string state as it evolves.  There is now vast literature on gravitational radiation from classical cosmic strings (see for example \cite{Damour:2000wa} \cite{Garfinkle:1988yi} \cite{Damour:2004kw}).  This is computed by Fourier transforming the functional derivative of the string action,
\begin{eqnarray}
\nonumber
 T^{\mu \nu}(k) &=&  \int d^D x \ e^{-ikx} \frac{\delta}{\delta g_{\mu \nu}} \frac{1}{2 \pi \alpha'} \int d^2 z \ g_{\mu \nu} \partial X^\mu {\bar \partial} X^\nu \delta^D (X(z, {\bar z}) - x)\\
 \label{gravsource}
&=& \frac{1}{2 \pi \alpha'} \int d^2 z \ \partial X^\mu {\bar \partial} X^\nu e^{-ikX} .
\end{eqnarray}
Note that this can be interpreted as a graviton vertex operator integrated over the worldsheet.  As explained previously, this is largest when both $k \cdot X_L(z)$ and $k \cdot X_R( {\bar z})$ develop saddle points (such as in a cusp), or when one does and the other develops a discontinuity (such as in a kink).  By substituting in a suitable solution for $X(z, {\bar z})$ around this point one can calculate the radiation emitted.  

We would now hope to calculate the same quantity for cosmic superstrings from first principles; that is, use vertex operators to represent the radiation (not necessarily gravitational) as well as for the cosmic string states themselves.  For the radiation emitted from a cosmic string of vertex operator $V_{\rm cs}(p)$, the amplitude is
\[ \mathcal A_{{\rm rad},f} (k) = \langle V_{{\rm cs}',f}(\infty;-p+k) V_{\rm rad} (1;-k) V_{\rm cs}(0;p) \rangle. \]
However, it would be much easier to compare to the classical answer if we could use an effective string action, at least for the massless radiation:
\[ S_{\rm eff} = \frac{1}{2 \pi \alpha'} \int d^2 z \left[ g_{\mu \nu} (X) \partial X^\mu {\bar \partial} X^\nu + \mathcal O (\alpha') \right] \]
where this is defined as performing the path integral over quantum fluctuations in a classical background,
\[ e^{-S_{\rm eff}[X_{\rm cl}]} = \int \left. [DX'] \right|_{X=X_{\rm cl}+X'} [Dh] e^{-S[X,h]} . \]
Evaluating this path integral is actually trivial because we know there should be no corrections to the classical action, string theory being {\it defined} by the condition that the worldsheet beta functionals are all zero!  Thus the classical source (\ref{gravsource}) is correct even for quantum superstrings (modulo the addition of fermions).  There will still be differences in the actual signal measured by a gravitational wave detector since the graviton must propagate according to the $\alpha'$-corrected General Relativity of string theory, but the source term will be the same.  Recently the small-scale structure which underlies this radiation emission has recently been studied in \cite{Polchinski:2006ee} \cite{Polchinski:2007rg} \cite{Dubath:2007wu}.

In order to ensure the beta functions are indeed zero it is necessary to represent the cosmic superstring as a physical string state (that is, satisfies the Virasoro constraints).  This is a different approach taken from \cite{Chialva:2006ak} who model the cosmic superstring as a semi-classical coherent state.  This then produces very small corrections in the source term, though still too small to be observed.

For massive radiated states it is completely straightforward to compute the emission decay rate using the same technique as that used for the reconnection probability, where the sum over amplitudes and final states can be computed from the distribution
\[ \sum_f \mathcal M_{{\rm rad},f}(k) V_{{\rm cs}',f}(0; p-k) =  C_{S_2} \left( \frac{\kappa}{2 \pi \sqrt{V}} \right)^2 \frac{1}{(2 \pi i)^2} \oint_0 d \epsilon d {\bar \epsilon} \ V_{\rm rad} (\epsilon,{\bar \epsilon}; -k) V_{\rm cs}(0;p) . \]
Of course the total decay rate must equal that given by the imaginary part of the one-loop mass correction,
\[ \Gamma = \frac{1}{m} {\rm Im} \ \langle V_{\rm cs}(-p) V_{\rm cs}(p) \rangle_{T^2} . \]
This is the approach used to calculate the decay rate for highly excited cosmic string `loops' which have broken off of large winding modes \cite{Chialva:2003hg} \cite{Chialva:2004ki}.  
%%%%%%%%%%%%%%%%%%%%%%%%%%%%%%%%%%%%%%%%%%%%%%%%%%%%%%%%%%%%%%%%%%%%%%%%%%%%%%%%
\section{Boundary Functionals and $(p,q)$ Strings}
%%%%%%%%%%%%%%%%%%%%%%%%%%%%%%%%%%%%%%%%%%%%%%%%%%%%%%%%%%%%%%%%%%%%%%%%%%%%%%%%
The methods we have developed here work exclusively for fundamental cosmic superstrings.  It would be theoretically and practically convenient to generalize this to D-strings \cite{Polchinski:1995mt} and $(p,q)$-strings \cite{Schwarz:1995dk}.  The interactions of these have been studied using either a string worldsheet with boundary conditions or as field theories living on the D-string worldvolume \cite{Jackson:2004zg} \cite{Hanany:2005bc} \cite{Hashimoto:2005hi}.

One might try to formulate such a general approach for string interactions by defining boundary functionals to represent the fundamental strings; that is, states defined on the unit circle which are a function of the coordinate modes $X_n = \frac{1}{2 \pi} \int_0 ^{2 \pi} X(\sigma) e^{-in \sigma} d \sigma$.  For the straight strings (formerly represented as wound tachyon vertex operators) these states $|W \rangle$ are simply gaussian distributions reflecting the fact that they are unexcited ground states,
\[ \langle X_n | W \rangle \propto e^{-\frac{n}{\alpha'} X_n X_{-n} }. \]
For the kinked fundamental string states defined in (\ref{kinkstate}) the answer is more interesting.  We attempt to construct such a state $| K \rangle$ by factoring the kink state as
\begin{equation}
\label{factored}
 | {\rm kinks} \rangle  = \Delta |K \rangle
 \end{equation}
 where $\Delta$ is the closed string propagator. To accomplish this, let us reexamine the earlier process used to construct the kink states in terms of a projection operator of $H$ and $P$.  Re-writing (\ref{holokink}) in terms of these operators, we get an expression which can be factorized into the desired form of (\ref{factored}):
\begin{eqnarray*}
 | {\rm kinks} \rangle  &=&  \frac{ \sin \pi P }{\pi P } \frac{ \sin (\pi H/2)}{\pi H/2} V_T(1;p_1) |p_2\rangle \\
 &=&  \Delta   \frac{\sin (\pi H/2)}{\alpha' \pi/2}  V_T(1;p_1) |p_2 \rangle .
\end{eqnarray*}
Thus we identify our boundary functional as
\begin{equation}
\label{bdef}
 |K \rangle =  \frac{4}{\pi \alpha' } \sin \frac{\pi H}{2} V_T(1;p_1) |p_2 \rangle .
 \end{equation}
Such a $\sin \pi H /2$ factor should be familiar from amplitudes which factor closed string amplitudes into (anti)holomorphic components.  To write this in a more useful form, represent the operator as the difference of exponentials
 \[ \sin \frac{\pi H}{2} = \frac{1}{2 i} \left( e^{i \pi H/2} - e^{- i \pi H/2} \right). \]
This can now easily act on $V_T(1;p_1)|p_2 \rangle$ to produce
\[ |K \rangle =  |+ \rangle - |- \rangle \]
where we have defined the coherent states
 \begin{eqnarray*} 
 | \pm \rangle &\equiv&  \frac{2}{\pi \alpha' i} :e^{ip_{1L}X_L{(i)}+ip_{1R}X_R(i)}: |p_2 \rangle \\
 &=& \frac{2(\pm i)^{ \frac{\alpha'}{2} (p_{1L} \cdot p_{2L}+p_{1R} \cdot p_{2R})} }{\pi \alpha' i} e^{\sqrt{ \frac{\alpha'}{2}} \sum_{n\geq1} p_{1L} \cdot \alpha_{-n}(\pm i)^n/n + p_{1R} \cdot {\tilde \alpha}_{-n}(\pm i)^n/n } |p_1+p_2 \rangle, \\
\langle X_n | \pm \rangle&\propto& \exp \left[ -\frac{n}{\alpha'} \left( X_n - i \alpha' p_{1R} (\pm i)^n/n \right) \left( X_{-n} - i \alpha' p_{1L} (\pm i)^n/n \right) \right].
 \end{eqnarray*}
Though there are an infinite number of oscillator excitations the propagator $\Delta$ will project only onto the on-shell portion of this.  Thus the state can be thought of heuristically as $V_T(p_2)$ at the origin and $V_T(p_1)$ placed at $\pm i$, though it is not a usual vertex operator $V(z, {\bar z})$ in the sense that ${\bar z}$ is not the complex conjugate of $z$.  

Even though we have defined these states $|W \rangle$ and $|K\rangle$ on the unit circle in an attempt to make them similar to D-branes, they are not boundary states in that they do not define boundary conditions: there is ``charge" (in the form of the vertex operator $V_T(p_1)$) located on the boundary so it is impossible to fix boundary conditions there.  It would be interesting to see whether this novel representation of the kink can be used for more efficient computation of the interactions.

%%%%%%%%%%%%%%%%%%%%%%%%%%%%%%%%%%%%%%%%%%%%%%%%%%%%%%%%%%%%%%%%%%%%%%%%%%%%%%%%
\section{Discussion and Conclusion}
%%%%%%%%%%%%%%%%%%%%%%%%%%%%%%%%%%%%%%%%%%%%%%%%%%%%%%%%%%%%%%%%%%%%%%%%%%%%%%%%
It is now possible to calculate the interaction and radiation for cosmic superstrings as well as their classical counterparts.  This is particularly exciting due to the increasing sensitivity of experiments to cosmic string radiation, stochastic as well as directed bursts.  Bounds on these currently exist from various sources such as LIGO S4, pulsar timing, big bang nucleosynthesis, and the cosmic microwave background, with potential further bounds from advanced LIGO and LISA  \cite{Siemens:2006vk} \cite{Wyman:2005tu} \cite{Siemens:2006yp} \cite{Hogan:2006va}.  These bounds are very dependent upon parameters such as the dimensionless string tension parameter $G \mu$, probability of reconnection $P$, loop size parameter $\epsilon$ and radiative parameter $\alpha$ \cite{Damour:2004kw}.  With the tools developed in this article one can begin to study the phenomenology of models based on superstring parameters $\alpha' $ and $g_s$, plus specifics of compactification \cite{Jackson:2006rb}. If even a single cosmic string is found, one can get an observational estimate of $P(\theta,v)$ from examining the kinks on the string as it has reconnected during its lifetime.  A low $P \sim g_s^2$ suggests a fundamental cosmic superstring such as those studied here, whereas a higher $P \sim 1$ suggests a classical vortex cosmic string or $(p,q)$-string.  The angle- and velocity-dependent reconnection probability can easily be incorporated into a simulation or analytic study \cite{Polchinski:2006ee} to study the abundance of cosmic strings present today.  It might also be relevant to string-gas studies since winding string interaction rates could influence the decompactification rate of the universe \cite{Brandenberger:1988aj}.  

%%%%%%%%%%%%%%%%%%%%%%%%%%%%%%%%%%%%%%%%%%%%%%%%%%%%%%%%%%%%%%%%%%%%%%%%%%%%%%%%
\section{Acknowledgments}
%%%%%%%%%%%%%%%%%%%%%%%%%%%%%%%%%%%%%%%%%%%%%%%%%%%%%%%%%%%%%%%%%%%%%%%%%%%%%%%%
It is a pleasure to thank T. Damour, P. Di Vecchia, K. Hashimoto, L. J. Huntsinger, S. S. Jackson, D. Kabat, O. Lunin, J. Lykken, V. Mandic, E. Martinec, and J. Polchinski for useful discussions, the organizers of the ``Cosmic Strings and Fundamental Strings" mini-workshop at Institut Henri Poincar\'{e} where this project was initiated, and especially N. Jones and G. Shiu who contributed much during the early stages of this project.  I am supported by NASA grant NAG 5-10842.
%%%%%%%%%%%%%%%%%%%%%%%%%%%%%%%%%%%%%%%%%%%%%%%%%%%%%%%%%%%%%%%%%%%%%%%%%%%%%%%%


\begin{thebibliography}{99}
%%%%%%%%%%%%%%%%%%%%%%%%%%%%%%%%%%%%%%%%%%%%%%%%%%%%%%%%%%%%%%%%%%%%%%%%%%%%%%%%

  %\cite{Witten:1985fp}
\bibitem{Witten:1985fp}
  E.~Witten,
  ``Cosmic Superstrings,''
  Phys.\ Lett.\ B {\bf 153}, 243 (1985).
  %%CITATION = PHLTA,B153,243;%%

%\cite{Copeland:2003bj}
\bibitem{Copeland:2003bj}
E.~J.~Copeland, R.~C.~Myers and J.~Polchinski,
``Cosmic F- and D-strings,''
JHEP {\bf 0406}, 013 (2004)
[arXiv:hep-th/0312067].
%%CITATION = HEP-TH 0312067;%%

%\cite{Becker:2005pv}
\bibitem{Becker:2005pv}
  K.~Becker, M.~Becker and A.~Krause,
  ``Heterotic cosmic strings,''
  Phys.\ Rev.\  D {\bf 74}, 045023 (2006)
  [arXiv:hep-th/0510066].
  %%CITATION = PHRVA,D74,045023;%%

  %\cite{Polchinski:2004ia}
\bibitem{Polchinski:2004ia}
  J.~Polchinski,
  ``Introduction to cosmic F- and D-strings,''
  arXiv:hep-th/0412244.
  %%CITATION = HEP-TH 0412244;%%
  
  %\cite{Davis:2005dd}
\bibitem{Davis:2005dd}
  A.~C.~Davis and T.~W.~B.~Kibble,
  ``Fundamental cosmic strings,''
  Contemp.\ Phys.\  {\bf 46}, 313 (2005)
  [arXiv:hep-th/0505050].
  %%CITATION = HEP-TH 0505050;%%
  
%\cite{Polchinski:1988cn}
\bibitem{Polchinski:1988cn}
J.~Polchinski,
``Collision Of Macroscopic Fundamental Strings,''
Phys.\ Lett.\ B {\bf 209}, 252 (1988).
%%CITATION = PHLTA,B209,252;%%

%\cite{Dai:1989cp}
\bibitem{Dai:1989cp}
  J.~Dai and J.~Polchinski,
  ``The Decay Of Macroscopic Fundamental Strings,''
  Phys.\ Lett.\ B {\bf 220}, 387 (1989).
  %%CITATION = PHLTA,B220,387;%%  
  
  \bibitem{Khuri:1993ax}
  R.~R.~Khuri,
  ``Veneziano amplitude for winding strings,''
  Phys.\ Rev.\ D {\bf 48}, 2823 (1993)
  [arXiv:hep-th/9303074];  R.~R.~Khuri,
  ``Classical dynamics of macroscopic strings,''
  Nucl.\ Phys.\ B {\bf 403}, 335 (1993)
  [arXiv:hep-th/9212029]; R.~R.~Khuri,
  ``Geodesic scattering of solitonic strings,''
  Phys.\ Lett.\ B {\bf 307}, 298 (1993)
  [arXiv:hep-th/9212026].
 %%CITATION = HEP-TH 9303074;%%
 %%CITATION = HEP-TH 9212029;%%
 %%CITATION = HEP-TH 9212026;%%
 
  %\cite{Mende:1994wf}
\bibitem{Mende:1994wf}
  P.~F.~Mende,
  ``High-energy string collisions in a compact space,''
  Phys.\ Lett.\ B {\bf 326}, 216 (1994)
  [arXiv:hep-th/9401126].
  %%CITATION = HEP-TH 9401126;%%
  
  %\cite{Jackson:2004zg}
\bibitem{Jackson:2004zg}
  M.~G.~Jackson, N.~T.~Jones and J.~Polchinski,
  ``Collisions of cosmic F- and D-strings,''
  JHEP {\bf 0510}, 013 (2005)
  [arXiv:hep-th/0405229].
  %%CITATION = HEP-TH 0405229;%%
  
%\cite{Jackson:2006rb}
\bibitem{Jackson:2006rb}
  M.~G.~Jackson,
  ``Cosmic superstring scattering in backgrounds,''
  JHEP {\bf 0609}, 071 (2006)
  [arXiv:hep-th/0608152].
  %%CITATION = HEP-TH 0608152;%%
 
   \bibitem{joebook}
 J. Polchinski, {\it String Theory Volumes I and II}, Cambridge: Cambridge University Press, 1998
 
%\cite{McLoughlin:2006tz}
\bibitem{McLoughlin:2006tz}
  T.~McLoughlin and X.~Wu,
  ``Kinky strings in $AdS(5) \times S^5$,''
  JHEP {\bf 0608}, 063 (2006)
  [arXiv:hep-th/0604193].
  %%CITATION = JHEPA,0608,063;%%
 
%\cite{Cornou:2006wx}
\bibitem{Cornou:2006wx}
  J.~L.~Cornou, E.~Pajer and R.~Sturani,
  ``Graviton production from D-string recombination and annihilation,''
  Nucl.\ Phys.\  B {\bf 756}, 16 (2006)
  [arXiv:hep-th/0606275].
  %%CITATION = NUPHA,B756,16;%%
  
 \bibitem{Melkumova:2006ua}
  E.~Y.~Melkumova, D.~V.~Gal'tsov and K.~Salehi,
  ``Dilaton and axion bremsstrahlung from collisions of cosmic
  (super)strings,''
  arXiv:hep-th/0612271.
  %%CITATION = HEP-TH 0612271;%%.  
  
 %\cite{Damour:2000wa}
\bibitem{Damour:2000wa}
T.~Damour and A.~Vilenkin,
``Gravitational wave bursts from cosmic strings,''
Phys.\ Rev.\ Lett.\  {\bf 85}, 3761 (2000)
[arXiv:gr-qc/0004075]; T.~Damour and A.~Vilenkin,
``Gravitational wave bursts from cusps and kinks on cosmic strings,''
Phys.\ Rev.\ D {\bf 64}, 064008 (2001)
[arXiv:gr-qc/0104026].
%%CITATION = GR-QC 0004075;%%
%%CITATION = GR-QC 0104026;%%
    
 \bibitem{mgj2}
 M.~G.~Jackson, work in progress.
    
 \bibitem{oneloop}
 M.~G.~Jackson, work in progress.
 
%\cite{Garfinkle:1988yi}
\bibitem{Garfinkle:1988yi}
  D.~Garfinkle and T.~Vachaspati,
  ``Fields due to Kinky, Cuspless, Cosmic Loops,''
  Phys.\ Rev.\ D {\bf 37}, 257 (1988).
  %%CITATION = PHRVA,D37,257;%%

%\cite{Damour:2004kw}
\bibitem{Damour:2004kw}
T.~Damour and A.~Vilenkin,
``Gravitational radiation from cosmic (super)strings: Bursts, stochastic
background, and observational windows,''
arXiv:hep-th/0410222.
%%CITATION = HEP-TH 0410222;%%

  %\cite{Polchinski:2006ee} 
\bibitem{Polchinski:2006ee}
  J.~Polchinski and J.~V.~Rocha,
  ``Analytic study of small scale structure on cosmic strings,''
  Phys.\ Rev.\  D {\bf 74}, 083504 (2006)
  [arXiv:hep-ph/0606205].
  %%CITATION = PHRVA,D74,083504;%%
  
  %\cite{Polchinski:2007rg}
\bibitem{Polchinski:2007rg}
  J.~Polchinski and J.~V.~Rocha,
  ``Cosmic string structure at the gravitational radiation scale,''
  arXiv:gr-qc/0702055.
  %%CITATION = GR-QC/0702055;%%
  
  %\cite{Dubath:2007wu}
\bibitem{Dubath:2007wu}
  F.~Dubath and J.~V.~Rocha,
  ``Periodic gravitational waves from small cosmic string loops,''
  arXiv:gr-qc/0703109.
  %%CITATION = GR-QC/0703109;%%
  
  %\cite{Chialva:2006ak}
\bibitem{Chialva:2006ak}
  D.~Chialva and T.~Damour,
  ``Quantum effects in gravitational wave signals from cuspy superstrings,''
  JCAP {\bf 0608}, 003 (2006)
  [arXiv:hep-th/0606226].
  %%CITATION = HEP-TH 0606226;%%
  
    %
\bibitem{Chialva:2003hg}
  D.~Chialva, R.~Iengo and J.~G.~Russo,
  ``Decay of long-lived massive closed superstring states: Exact results,''
  JHEP {\bf 0312}, 014 (2003)
  [arXiv:hep-th/0310283].
  %%CITATION = HEP-TH 0310283;%%
  
  %
\bibitem{Chialva:2004ki}
  D.~Chialva and R.~Iengo,
  ``Long lived large type II strings: Decay within compactification,''
  JHEP {\bf 0407}, 054 (2004)
  [arXiv:hep-th/0406271].
  %%CITATION = HEP-TH 0406271;%%
   
%\cite{Polchinski:1995mt}
\bibitem{Polchinski:1995mt}
  J.~Polchinski,
  ``Dirichlet-Branes and Ramond-Ramond Charges,''
  Phys.\ Rev.\ Lett.\  {\bf 75}, 4724 (1995)
  [arXiv:hep-th/9510017].
  %%CITATION = PRLTA,75,4724;%%
  
  %\cite{Schwarz:1995dk}
\bibitem{Schwarz:1995dk}
  J.~H.~Schwarz,
 ``An SL(2,Z) multiplet of type IIB superstrings,''
  Phys.\ Lett.\  B {\bf 360}, 13 (1995)
  [Erratum-ibid.\  B {\bf 364}, 252 (1995)]
  [arXiv:hep-th/9508143].
  %%CITATION = PHLTA,B360,13;%%

  %\cite{Hanany:2005bc}
\bibitem{Hanany:2005bc}
  A.~Hanany and K.~Hashimoto,
  ``Reconnection of colliding cosmic strings,''
  JHEP {\bf 0506}, 021 (2005)
  [arXiv:hep-th/0501031].
  %%CITATION = JHEPA,0506,021;%%
  
  %\cite{Hashimoto:2005hi}
\bibitem{Hashimoto:2005hi}
  K.~Hashimoto and D.~Tong,
  ``Reconnection of non-abelian cosmic strings,''
  JCAP {\bf 0509}, 004 (2005)
  [arXiv:hep-th/0506022].
  %%CITATION = JCAPA,0509,004;%%
  
%\cite{Siemens:2006vk}
\bibitem{Siemens:2006vk}
  X.~Siemens, J.~Creighton, I.~Maor, S.~Ray Majumder, K.~Cannon and J.~Read,
  ``Gravitational wave bursts from cosmic (super)strings: Quantitative analysis and constraints,''
  Phys.\ Rev.\  D {\bf 73}, 105001 (2006)
  [arXiv:gr-qc/0603115].
  %%CITATION = PHRVA,D73,105001;%%
      
  %\cite{Wyman:2005tu}
\bibitem{Wyman:2005tu}
  M.~Wyman, L.~Pogosian and I.~Wasserman,
  ``Bounds on cosmic strings from WMAP and SDSS,''
  Phys.\ Rev.\ D {\bf 72}, 023513 (2005)
  [Erratum-ibid.\ D {\bf 73}, 089905 (2006)]
  [arXiv:astro-ph/0503364];
   N.~Bevis, M.~Hindmarsh, M.~Kunz and J.~Urrestilla,
  ``Fitting CMB data with cosmic strings and inflation,''
  arXiv:astro-ph/0702223;
    N.~Bevis, M.~Hindmarsh, M.~Kunz and J.~Urrestilla,
 ``CMB polarization power spectra contributions from a network of cosmic strings,''
  arXiv:0704.3800 [astro-ph].
      %%CITATION = ASTRO-PH 0503364;%%
  %%CITATION = ARXIV:0704.3800;%
  %%CITATION = ASTRO-PH/0702223;%%
  
%\cite{Siemens:2006yp}
\bibitem{Siemens:2006yp}
  X.~Siemens, V.~Mandic and J.~Creighton,
  ``Gravitational wave stochastic background from cosmic (super)strings,''
  Phys.\ Rev.\ Lett.\  {\bf 98}, 111101 (2007)
  [arXiv:astro-ph/0610920].
  %%CITATION = PRLTA,98,111101;%%
  
%\cite{Hogan:2006va}
\bibitem{Hogan:2006va}
  C.~J.~Hogan,
  ``Gravitational wave sources from new physics,''
  AIP Conf.\ Proc.\  {\bf 873}, 30 (2006)
  [arXiv:astro-ph/0608567].
  %%CITATION = APCPC,873,30;%%

    %
\bibitem{Brandenberger:1988aj}
  R.~H.~Brandenberger and C.~Vafa,
  ``Superstrings in the Early Universe,''
  Nucl.\ Phys.\  B {\bf 316}, 391 (1989);  R.~Easther, B.~R.~Greene and M.~G.~Jackson,
  ``Cosmological string gas on orbifolds,''
  Phys.\ Rev.\ D {\bf 66}, 023502 (2002)
  [arXiv:hep-th/0204099];   R.~Easther, B.~R.~Greene, M.~G.~Jackson and D.~Kabat,
  ``String windings in the early universe,''
  JCAP {\bf 0502}, 009 (2005)
  [arXiv:hep-th/0409121].
    %%CITATION = NUPHA,B316,391;%%
  %%CITATION = HEP-TH 0204099;%% 
    %%CITATION = HEP-TH 0409121;%%
    
 \end{thebibliography}
\end{document}